\documentclass[11pt, letterpaper]{article}
\usepackage[pdftex]{graphicx}
\usepackage{txfonts}
\usepackage{authblk}
\usepackage{cite}
\usepackage{tabularx}

\begin{document}
\title{Quantifying moral foundations from various topics on Twitter conversations}
\author[1]{\small Rishemjit Kaur}
\author[2,3]{\small Kazutoshi Sasahara\thanks{Correspondence should be addressed to K.S. (sasahara@nagoya-u.jp)}}
\affil[1]{\small CSIR-Central Scientific Instruments Organisation, Chandigarh, India}
\affil[2]{\small Department of Complex Systems Science, Nagoya University, Nagoya, Japan}
\affil[3]{\small Center for Complex Networks and Systems Research, Indiana University, Bloomington, IN, USA}
\date{}

\maketitle

\begin{abstract}
Moral foundations theory explains variations in moral behavior using innate moral foundations: Care, Fairness, Ingroup, Authority, and Purity, along with experimental supports.  
However, little is known about the roles of and relationships between those foundations in everyday moral situations. 
To address these, we quantify moral foundations from a large amount of online conversations (tweets) about moral topics on the social media site Twitter. 
We measure moral loadings using latent semantic analysis of tweets related to topics on abortion, homosexuality, immigration, religion, and immorality in general, showing how the five moral foundations function in spontaneous conversations about moral violating situations. 
The results indicate that although the five foundations are mutually related, Purity is the most distinctive foundation and Care is the most dominant foundation in everyday conversations on immorality. 
Our study shows a new possibility of natural language processing and social big data for moral psychology.
\end{abstract}

\thispagestyle{empty}
\newpage

\section{Introduction}

Social media communication is a regular part of our daily life and a large amount of linguistic information is posted and digitally recorded every day. 
Our digital behavioral traces allow us to quantify human behavior in a natural setting to complement experimental data. 
The availability of social big data for social science research is now widely acknowledged \cite{Lazer:2009ci,Golder:2014jm}, and computational social science provides new insights into human nature in the digital era \cite{Sasahara:2013eu,Takeichi:2015ia,Sasahara2016}. 

In this paper, we study moral behavior using social big data from a popular microblogging platform, Twitter. 
One of the most influential theories about moral behavior is Moral Foundations Theory (hereafter, MFT) proposed by Haidt and Joseph \cite{Haidt2012}. MFT can explain variations in moral behavior on the basis of the following innate moral foundations: 

\begin{itemize}
\item Care: disliking the pain of others and feeling of protecting the vulnerable;
\item Fairness: doing the right thing or justice based on shared rules;
\item Ingroup: being loyal to social groups, including family and nation;
\item Authority: respecting and obeying tradition and legitimating authority;
\item Purity: feeling an antipathy for disgusting things and contamination.
\end{itemize}

According to Haidt et al.~\cite{Haidt2008}, Care and Fairness, which focus on individuals' rights and freedom, are grouped together and referred to as `individualizing foundations,' whereas Ingroup, Authority, and Purity are mainly centered around binding people together for the welfare of community or group and are called `binding foundations.' 
MFT is especially successful in explaining political ideology and cross-cultural differences in moral behavior. 
For example, previous studies have shown that political progressives or `liberals' stress only Care and Fairness in moral judgment, whereas political conservatives equally stress all moral foundations \cite{Haidt2012}. 
Another study has demonstrated that people in collectivist societies are more sensitive to violations of the community-related moral foundations, whereas people in Western societies are more likely to discriminate between care-related violations and convention-related violations \cite{Haidt2012}. 
While moral foundations have been confirmed experimentally, more recent studies have used natural language processing (NLP) techniques and social big data \cite{Sagi2014,Sagi:2014un,Dehghani2016}: they measured the moral loadings from written expressions on a particular real-world event (2013 government shutdown in the US) from Twitter, showing that Purity is related to social distance. 
The findings were then validated experimentally. 
These studies have led to a new method with empirical evidence for MFT, yet we know little about the roles and relationships between moral foundations in everyday moral situations.

To address these issues, we quantified moral foundations from various topics in everyday conversations on Twitter. 
More specifically, we examined the relationships of five moral foundations by measuring moral loadings from a large amount of posted messages or tweets that are related to moral violating situations. 
Moreover, we analyzed tweets about moral topics such as abortion, homosexuality, immigration, and religion for insights into the roles of moral foundations in various topics.
\section{Method}

\subsection{Data collection}
We collected tweets related to moral concerns---abortion (`abortion'), homosexuality (`homosexuality' OR `homosexual'), immigration (`immigration' OR `immigrant'), religion (`religion' OR `religious'), and immorality (`immorality' OR `immoral') that were posted between March 1 and April 24, 2016 using the Twitter Search API (https://dev.twitter.com/docs/api/). 
The search queries written in brackets were selected to observe daily conversations on moral topics. 
English tweets were used for our text analysis.
The sizes of the datasets are 1516119 tweets for abortion, 456674 tweets for homosexuality, 2102886 tweets for immigration, 4628102 tweets for religion, and 217975 tweets for immorality, respectively. 
The topic datasets (i.e., tweets with `abortion', `homosexuality', `immigration', or `religion') were used only to identify keywords and context words associated with these topics. 
Except these, our analysis has been done with the immorality dataset (i.e., tweets with `immorality').

\subsection{Data pre-processing}
Each tweet consists of multiple fields such as text, retweeted status, language, and other meta-data. 
We extracted `text' field and `retweeted\_status.text' field if the tweet has been re-tweeted.
To clean the texts, we removed stop words (e.g., `a', `an', `the') that are defined in NLTK library (http://www.nltk.org), URLs, screen names (e.g., @BarackObama), special characters (e.g., ! and \$), numbers, leading and trailing white-spaces, and short words with length less than three (e.g., th, gf) from tweets. 
After that, hashtag symbol was removed from tweets (e.g., `\#harm' was replaced with `harm'). 
Furthermore, the words used in queries for Twitter Search API were removed because we are interested in popular words that characterize moral topics except those used in queries. 
For example, in the case of the immorality dataset, `immoral’ and `immorality’ were removed and thus not considered as keywords and context words for subsequent analyses. 
Further, we removed the duplicate tweets from our dataset. 
Finally, we tokenized tweet texts by splitting on white-spaces. Suppose a tweet ``@CharlesMBlow 50\% marginal taxrates aren't immoral. Letting the majority of public school kids live in poverty is. https://t.co/grEgeu9lPj”, by following the steps stated above we are left with `marginal', `taxrates', `letting', `majority', `public', `school', `kids', `live', `poverty.' 
This pre-processing was applied to the topic datasets and the immorality dataset after which we refer to them as the topic corpora and the immorality corpus, respectively. 

\subsection{Quantification of moral foundations}
To quantify the moral foundations from the Twitter corpora, we used a method proposed by Dehghani et al.~\cite{Dehghani2016} with several modifications. 
Simply speaking, the method is based on the latent semantic analysis (LSA)~\cite{Deerwester1990, Landauer1997}: using a bag-of-words model, a corpus is represented by a word-context matrix and by reducing its dimensionality we get low-dimensional word vectors, in which similar meaning words are represented by similar vectors. 
With this method, we can measure moral foundations from everyday tweets by comparing the words in the moral foundations dictionary (described below) and words in the Twitter corpora. 
The details are described below.

\subsubsection{Moral foundations dictionary}
The moral foundations dictionary (hereafter, the MF dictionary) was created by Graham and Haidt~\cite{Graham2009}. It is available online at http://moralfoundations.org. 
The MF dictionary lists the words and word stems associated with five moral foundations---Care (called `Harm' in the MF dictionary), Fairness, Ingroup, Authority and Purity, along with general words associated with morality and immorality. 
These are further divided into two categories, `virtue' and `vice.' 
Virtue words are foundation-supporting words (e.g. safe* and shield for Care virtue), whereas vice words are foundation-violating words (e.g. kill and ravage for Care vice). 
To limit our analysis of moral violating situations, we used 149 vice words in the MF dictionary.

\subsubsection{Selection of keywords and context words}
\begin{figure}[!t]
\centering
\includegraphics[width=\textwidth]{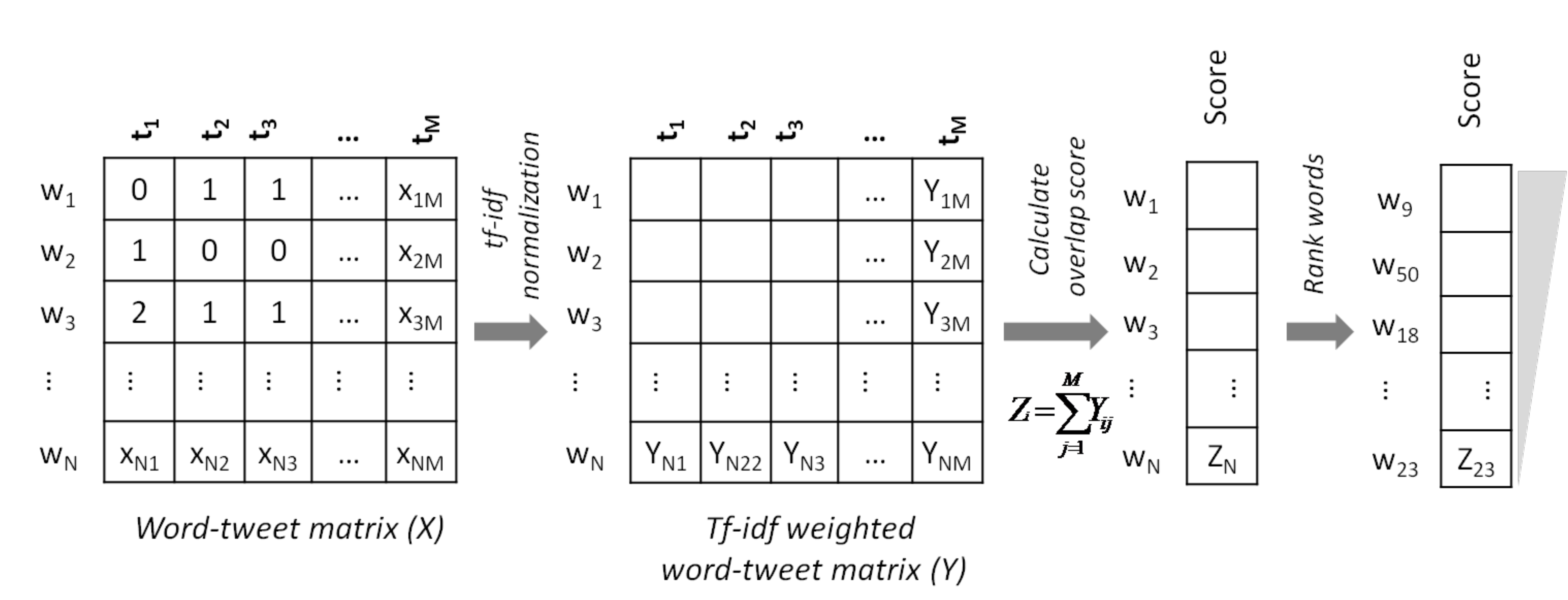}
\caption{Selection of keywords and context words}
\label{fig:selection}
\end{figure}

To construct a word-context matrix used for the succeeding analysis, we selected keywords and context words based on the tf-idf score from the topic corpora and the immorality corpus (Fig.~\ref{fig:selection}). 
From each of the Twitter corpora, we created a word-tweet matrix $X$, respectively, in which each row denotes word $w$ and each column denotes tweet $t$, and the element $X_{ij}$ denotes the frequency of word $w_i$ in tweet $t_j$. 
Then, this matrix was converted to a tf-idf weighted matrix $Y$ by the following equation:
\begin{equation}
\label{eq:tfidf}
     tf{\mathchar`-}idf(w_i,t_j) = tf(w_i ,t_j)(\log(M+1) - \log(df(w_i))),  
\end{equation}
where $tf(w_i , t_j)$ is the frequency of word $w_i$ in tweet $t_j$, $M$ is the total number of tweets and $df(w_i)$ is the number of tweets containing word $w_i$. 
The overlap score of a word $w_i$ in $Y$ is computed as a measure of word importance~\cite{Manning:2008:IIR:1394399}: 

\begin{equation}
\label{eq:score}
    Score(w_i) = \sum_{j=0}^{M} (tf{\mathchar`-}idf(w_i, t_j)).
\end{equation}
According to (\ref{eq:score}), the words were ranked in the decreasing order and the top $N_1$ words were selected as keywords and the top $N_2$ words as context words for a word-context matrix in order to use the subsequent analysis. 
Using the same settings of Dehgani et al.~\cite{Dehghani2016}, we set $N_1=2000$ and $N_2=20000$.

\subsubsection{Word-context matrix and singular value decomposition}
In using keywords as columns and context words as rows, we created a word-context matrix $C$, in which the element $C_{ij}$ represents the number of co-occurrence of word $w_i$ and context word $w_j$. 
In contrast to Dehghani et al. \cite{Dehghani2016}, we further converted it to the positive pointwise mutual information (PPMI) based matrix so as to not assign a higher weight to a popular general word that is irrelevant to moral issues \cite{Jurafsky2014}:
\begin{equation}
\label{eq:PPMI}
    PPMI(v_i, v_j) = max(\log_2 (P(v_i, v_j) / P(v_i)P(v_j)) ,0), 
\end{equation}
Here, $v_i$ and $v_j$ are the words in $C$, $P(v_i)$ is the occurrence probability of a word $v_i$ and $P(v_i, v_j)$ is the joint occurrence probability of $v_i$ and $v_j$. 
Using the PPMI-based word-context matrix $C$, we applied singular value decomposition (SVD) to achieve a lower dimensional representation of keywords: $C = U\sum V^*$ where $U$ is a left singular matrix, $\sum$ is a diagonal matrix and $V^*$ is the right singular matrix. 
The top $k$ dimensions (in our case $k = 100$) of $U$ matrix are retained, and each row of this reduced matrix represents a word in $k$ dimensions. The matrices $\sum$ and $V^*$ are discarded. 
In this way, SVD converts the high-dimensional and sparse word-context matrix into a lower dimensional, real-valued matrix, which represents the semantic relationships between words~\cite{Jurafsky2014}. 

\subsubsection{Construction of context vectors}
The resulting word vector space is linear. Therefore, it is possible to approximate a text by adding corresponding word vectors~\cite{Dehghani2016}. 
In doing this, we constructed tweet context vectors, topic context vectors, and moral foundation (MF) context vectors. 

\begin{figure}[hb]
\centering
\includegraphics[width=3in]{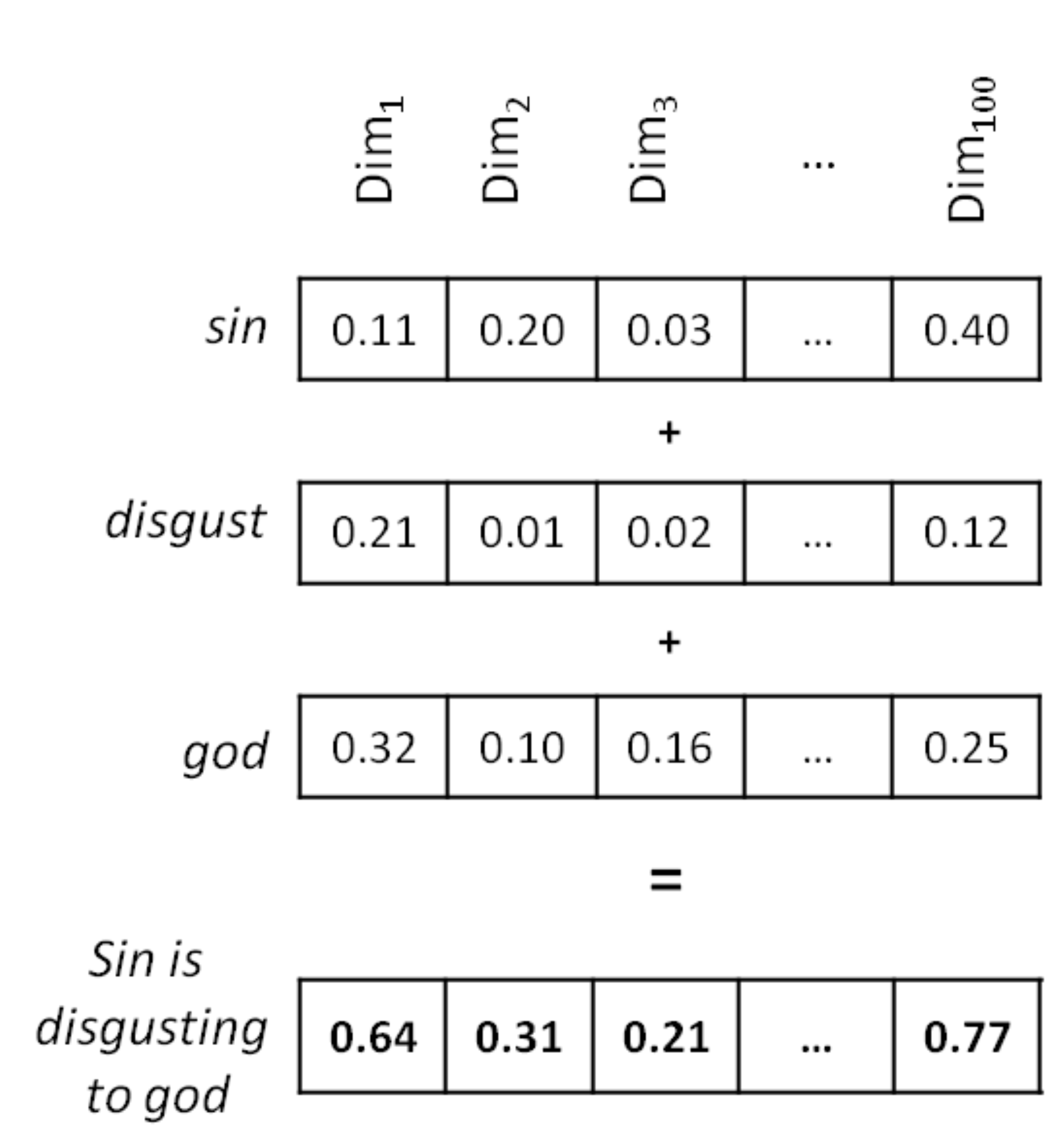}
\caption{Example of the construction of a tweet context vector}
\label{fig:CV}
\end{figure}

Fig.~\ref{fig:CV} shows an example of how to construct a tweet context vector. 
Given a tweet, `Sin is disgusting to god.’ the tweet context vector is obtained by the addition of the vectors of corresponding words `sin’, `disgust’, and `god’. Note that the stop words `is’ and `to' were removed during pre-processing. 
Similarly, we constructed MF context vectors by adding the vectors corresponding to the MF dictionary's words present in the immorality corpus, and topic context vectors by adding the vectors corresponding to topic specific words. 
As mentioned earlier, we have used the topic corpus to select keywords describing a specific topic (e.g., topic `religion' may be described by various words such as `God', `church', `religious', `Islam', `Christianity', `Hindu'). 
For comparison, we created the topic context vectors using 10 keywords and 100 keywords, respectively.

\subsubsection{Measurement of moral loadings}
\begin{figure}[!t]
\centering
\includegraphics[width=\textwidth]{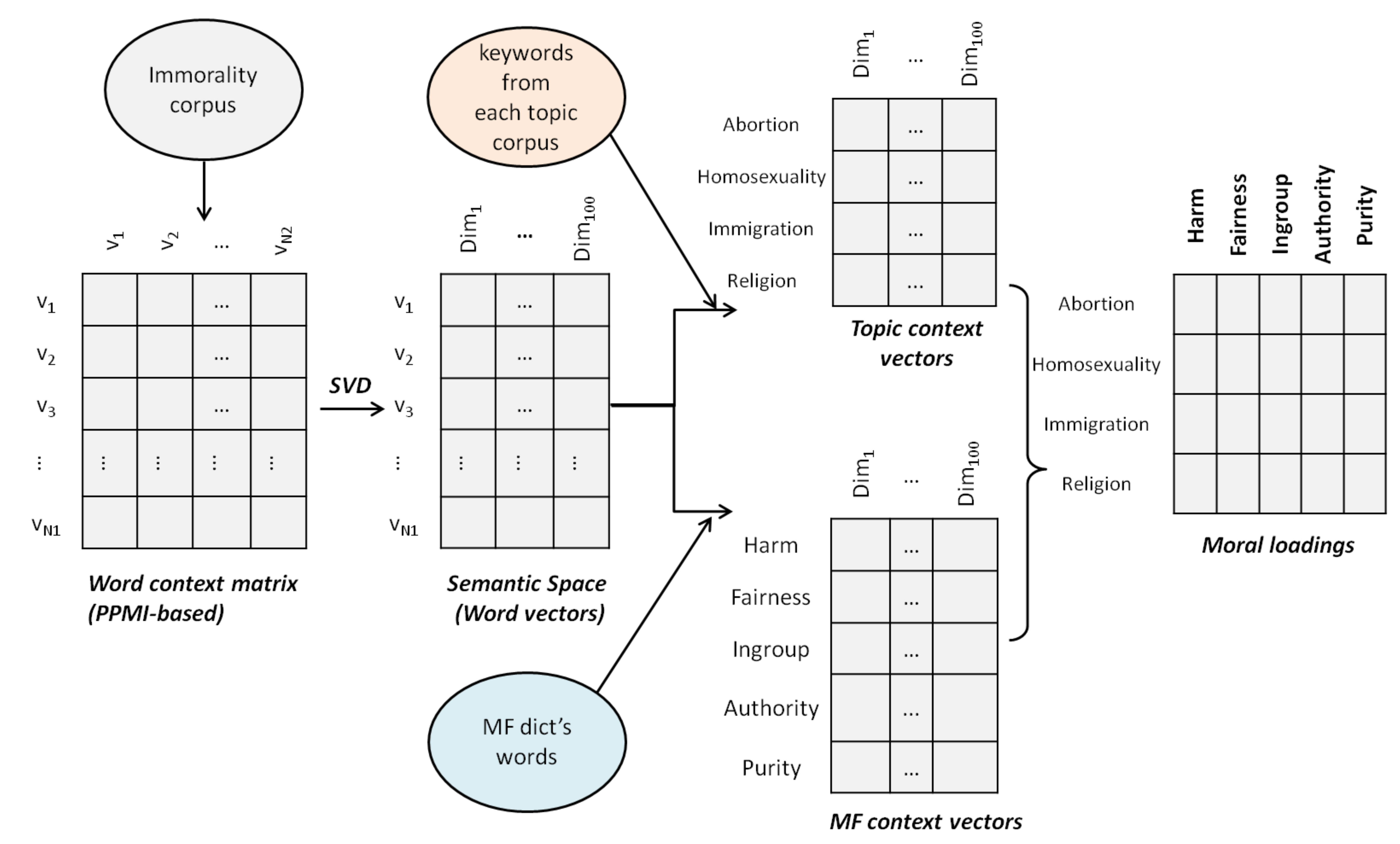}
\caption{Measurement of moral loadings for topics}
\label{fig:moralloadings}
\end{figure}
Of the three kinds of context vectors described above, one possible way to quantify moral foundations from texts is to measure moral loadings \cite{Dehghani2016}, which is defined as similarity to MF context vectors. 
For example, the moral loading of a tweet is measured by the cosine similarity between the tweet context vector and the MF context vector. 
The cosine value of 1 signifies synonymy of expressions, while the cosine value of 0 indicates that these are semantically unrelated. In this manner, if tweet context vectors (TV) are represented by $\langle TV_{1}, TV_2, TV_3,...,TV_M \rangle$ and MF context vectors (MV) by $\langle MV_{\rm Care}, MV_{\rm Fairness}, MV_{\rm Ingroup}, MV_{\rm Authority}, MV_{\rm Purity}\rangle$, then the moral loadings are summarized in a $M \times 5$ matrix whose element represents the similarity between a tweet and each moral foundation. 
Similarly, we can compute moral loadings for topics (i.e., abortion, homosexuality, immigration, religion), which results in a 4 (number of topics) $\times$ 5 (number of moral foundations) matrix representing similarity between topics and moral foundations. 
This procedure is illustrated in Fig.~\ref{fig:moralloadings}.
\section{Results}
\subsection{Immorality in everyday tweets}
\begin{figure}[!th]
\centering
\includegraphics[width=3.5in]{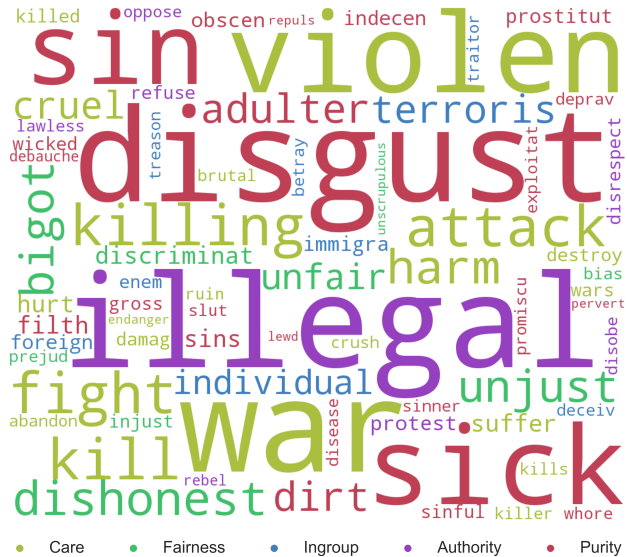}
\caption{Word cloud of the MF dictionary’s vice words present in the immorality corpus}
\label{fig:wordcloud}
\end{figure}

We analyzed the immorality corpus in terms of moral violating situations and found that 81.1\% of the MF dictionary’s vice words were present in the corpus. 
Fig.~\ref{fig:wordcloud} shows examples of the MF dictionary’s vice words present in the corpus, in which the size of a word is proportional to its occurrence frequency and colors represent different moral foundations.
For example, words such as `war', `killing', `violen*', `attack' are most frequently occurring words from Care foundation, and words such as `sin', `sick', `disgust', `dirt', `adulter*' were present in the Purity foundation.
Note that the MF dictionary’s words such as `spurn', `favoritism', `jilt*', `obstruct', `blemish' are not shown in this figure because either these words did not pass the keyword selection described before or these were not present in our corpus.
It is worth noticing that `Illegal' is the most frequent word in the five foundations.
This suggests that illegal issues would be the most immoral in everyday moral situations.

Table \ref{moralLoadings} shows examples of the tweets with the highest moral loadings from each moral foundation. 
Tweet \#1 has the highest similarity with the MF context vector for Care. 
This could be because of the presence of the word `kills' in the tweet. 
Tweet \#4 is an interesting example where the highest correlation of 0.553 is obtained with Authority, the second highest of 0.348 was with Ingroup, and the third highest was 0.346 and with Fairness. 
If we look at the MF dictionary, the word `treasonous' belongs to both Ingroup and Authority, and the word `illegal' belongs to Authority. Thus, this tweet has a higher correlation with Authority than Ingroup. 
This tweet also has a high correlation with Fairness even though it does not include any word from the Fairness category in the MF dictionary. 
It is because of the presence of words such as `patriotic' (0.33), `government' (0.23), etc. in the tweet which are highly correlated with Fairness in terms of cosine similarity.
This is an example case where a standard word counting approach~\cite{Graham2009,Clifford2013} may fail because the MF dictionary includes only a limited number of words (e.g., 149 words in the vice category), and other everyday moral words are not considered.
In contrast, the matrix $U$ resulting from the moral loading method provides an extended version of the MF dictionary, which we will detail later. 
Hence, it may be applicable to a wider class of texts.

\begin{table*}
\caption{Tweets showing the highest moral loadings with each foundation}
\label{moralLoadings}
\centering
\begin{tabularx}{\linewidth}{|c|X|c|c|c|c|c|}\hline
   \# & Tweet & Care  & Fairness & Ingroup & Authority & Purity \\\hline
    1  & @harikriss it should it's immoral and kills innocent people 
     & \textbf{0.772} & -0.016   & 0.199   & 0.063     & 0.113  \\\hline
    2  & @goat777face @ColinUlster96 @dkm49321 Yeah, not being to discriminate is so unbelievably immoral and unjust.                          & 0.184 & \textbf{0.712}    & 0.177   & 0.311     & 0.006  \\\hline
    3  & @DanWosHere @malcolmtyson @JulianBurnside Foul immoral traitors will go to any lengths to defend  the enemy they've allied with.      & 0.225 & 0.189    & \textbf{0.544}   & 0.336     & 0.186  \\\hline
    4  & @Scribbles646 @Snowden So, we agree arming Al Qaeda is treasonous. Exposing the illegal/immoral actions of a government is patriotic. & 0.218 & 0.346    & 0.348   & \textbf{0.553}     & 0.007  \\\hline
    5  & \#PresstitutesDay Sick of these indecent, perverted, shameless, wicked, sinful, immoral, lewd, self-indulgent anti nationals.           & 0.134 & 0.249    & 0.238   & 0.044     & \textbf{0.684}  \\\hline
    
\end{tabularx}
\end{table*}

To examine which moral foundations are actually used as well as how often those are used in Twitter conversations, we calculated moral loadings for each tweet and classified them into one moral foundation with maximum moral loading value, i.e., the tweet was assigned a dominant moral foundation with which it had the highest similarity. 
For example, tweet \#1 in Table \ref{moralLoadings} was assigned to Care. 
We calculated the total number of tweets in each moral foundation. 
This result is summarized in Table \ref{numberTweets}. 
In the immorality corpus, we observed that the maximum number of tweets (21135) belonged to Care, and the minimum number of tweets (4932) belonged to Authority. 
This suggests that Care is the most dominant foundation that people are knowingly or unknowingly concerned with when discussing in the context of immorality, and that people would less pay attention to Authority violation in everyday moral situations. 

\begin{table}
\caption{Number of tweets present in each moral foundation}
\label{numberTweets}
\centering
\begin{tabular}{|c|c|}\hline
    Foundation  & Number of tweets\\\hline
    Care & 21135\\\hline
    Fairness & 15731\\\hline
    Ingroup & 6665\\\hline
    Authority & 4932 \\\hline
    Purity &  14587 \\\hline
\end{tabular}
\end{table}
   
Table \ref{cosMF} shows the relationships between the five moral foundations in the context of moral violating situations.
Here, the cosine similarity should be close to zero if the two foundations are independent or orthogonal to each other. 
We do observe a high correlation between Ingroup and Authority (0.598), which seems to be natural because both belong to the binding foundations~\cite{Haidt2008}. 
We expected the correlation between Care and Fairness to be higher than correlation between Care and Ingroup because both are individualizing moral foundations. However, this was not observed. 
Table \ref{cosMF} also shows that Purity is one of the foundations that does not have compelling correlations with either individualizing or binding foundations. 

\begin{table}
\caption{Cosine similarities between MF context vectors}
\label{cosMF}
\centering
\begin{tabular}{|c|c|c|c|c|c|}\hline

    Foundation & Care & Fairness & Ingroup & Authority & Purity \\ \hline
    Care       & -    & 0.113    & \textbf{0.394}   & 0.133     & 0.229  \\ \hline
    Fairness   & -    & -        & 0.147   & 0.223     & 0.14   \\ \hline
    Ingroup    & -    & -        & -       & \textbf{0.598}     & 0.239  \\ \hline
    Authority  & -    & -        & -       & -         & 0.081  \\ \hline
    Purity     & -    & -        & -       & -         & -  \\ \hline
    
\end{tabular}
\end{table}

\subsection{Extended MF dictionary}
Using the immorality corpus, we extended the MF dictionary by adding the semantically related words using the resulting word-context matrix. 
That is, we again calculated cosine similarities between five MF context vectors and word vectors and then listed the top 100 most similar words for each foundation. 
In this way, the extended MF dictionary had 500 total words. 

We compared the original and extended dictionaries and found several similarities and dissimilarities. 
While most words are commonly present in the same foundation category (e.g., `kill', `kills', `attack*' are involved in Care), others present in different foundation categories (e.g., `insurgent' is in Authority in the original dictionary but it appears in Care in the extended dictionary). 
Some words appear in more than one foundation in the extended dictionary (e.g., `terroris*', which is present in Ingroup in the original dictionary, appears both in Care and Ingroup). 
Importantly, words not present in the original dictionary have been added in the extended dictionary (e.g., `bombing', `isis', `torture', `murder', `nazis' , `stereotypes' in Care, `racist', `elitists', `racisim' in Ingroup and `lust', `flesh', `devil' in Purity). 
Furthermore, some morally neutral words, e.g., `Kashmir' (the name of a place in India), is present in Care, which in recent years has seen political and social instability leading to violent clashes and a loss of life and property. 
The presence of words such as 'homosexual' and `lesbians' along with other words of Purity (e.g., `disgust', `sinful', `sin') reflects the attitude of people towards these topics similar to `abortion' in Care. 
In this way, the extended MF dictionary reflects actuarial word usages in everyday conversations.

\begin{figure}[th]
\centering
\includegraphics[width=4in]{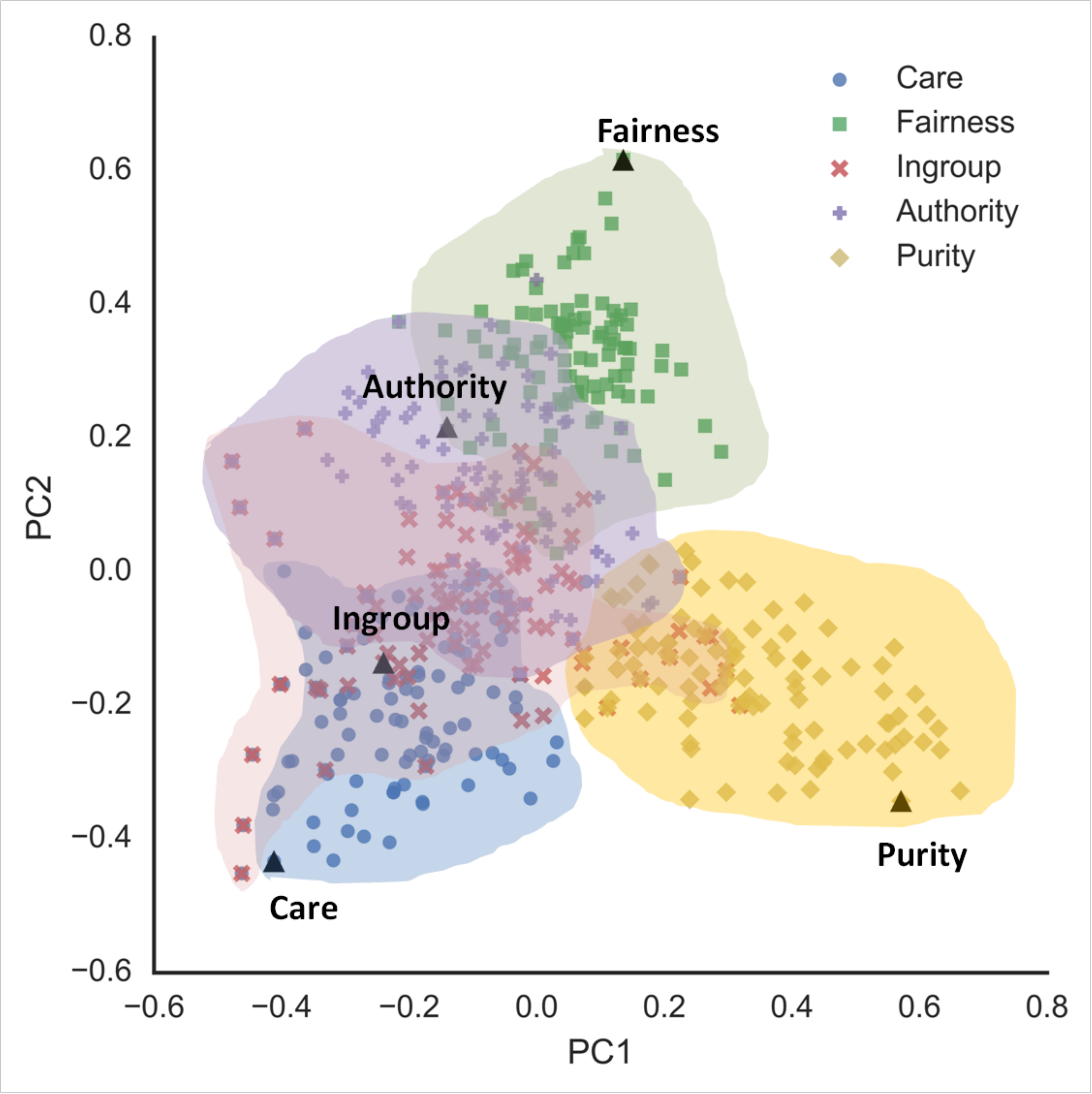}
\caption{PCA plot of the extended MF dictionary's words along with five moral context vectors}
\label{fig:PCAplot}
\end{figure}

To further examine how these foundations relate to each other, we mapped the extended dictionary's word vectors along with moral foundations vectors on a 2D plane using principal component analysis (PCA)~\cite{Jolliffe2002}. 
Fig.~\ref{fig:PCAplot} shows that except Purity, the words belonging to other four foundations overlap with each other along the PC1 axis. 
Thus, the PC1 axis differentiates Purity and non-Purity foundations, which can be explained in terms of `person-based attributes' vs.`situation-based attributes'~\cite{Chakroff:2015if}.
The other four foundations are differentiated along the PC2 axis, although with some overlaps. 
Care and Fairness are placed at the extreme ends along the PC2 axis, indicating that these two foundations are dissimilar to each other. This is similar to the result found in the previous section.
This result indicates that Care and Fairness are both individualizing tendencies but function in different manners. 
Overall, both axes well differentiate Purity, Fairness, and Care from each other, but there is a significant overlap between Authority and Ingroup. 
It can also be seen that Ingroup has an overlap with Care, and Authority overlaps with Fairness to some extent. 
These results imply that the five moral foundations are not mutually exclusive---at least in everyday conversations on immorality.

\subsection{Roles of moral foundations in various topics}
To investigate the roles of moral foundations on different topics, we measured the moral loadings between different topics and moral foundations. Table~\ref{topic10} shows the result when we used the top 10 keywords according to score (eq. (\ref{eq:score})) for topic context vectors. 
We see that topics `abortion’ and `religion’ are correlated to Care, whereas topics `immigration’ and `homosexuality’ have the highest similarities with Ingroup and Purity, respectively. 
The similarity of homosexuality with Purity can be explained as it has been found by Pizarro et al.~\cite{Pizarro2011} that purity violations evoke feeling of disgust. Disgust is also considered to be positively correlated with the negative attitudes towards homosexuals~\cite{Terrizzi2010,Smith2011}.  
Graham et al.~\cite{Graham2012} mentioned that immigrants are a trigger for Purity violations, but our results show that its highest correlation is with Ingroup violations. 

The differences between moral concerns of Republicans and Democrats about `abortion’ were discussed by Sagi et al.~\cite{Sagi2014}. 
Their results showed that Democrats were mostly concerned with Fairness, whereas Republicans were concerned with the Purity aspects of abortion. In our case, abortion has the highest correlation with Care. 

When we used the top 100 keywords according to score (eq. (\ref{eq:score})) for constructing topic context vectors from topic corpora, all topics were maximally correlated with Care (Table~\ref{topic100}). 
This result suggests that although the most popular keywords (e.g, top 10 words) are related to the corresponding topics, many words ranked below the top 10 were related to Care. 
This is partly supported by the fact that the maximum number of tweets in our dataset were predominantly related to Care as shown Table~\ref{numberTweets}.

 \begin{table}[!th]
 \caption{Moral loadings for topics (based on top 10 keywords)}
 \label{topic10}
 \centering
 \begin{tabular}{c}
 \includegraphics[width=\linewidth, clip=true, trim=1.75cm 25cm 6.5cm 1.85cm]{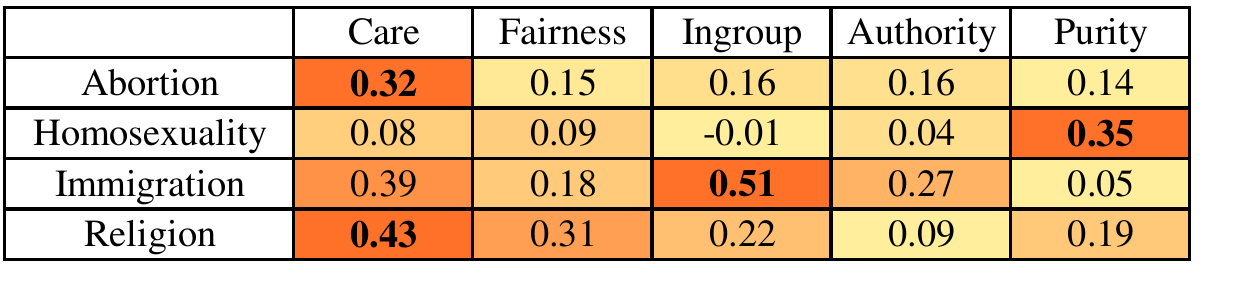} 
 \end{tabular}
 \end{table}

 \begin{table}[ht]
 \caption{Moral loadings for topics (based on top 100 keywords)}
 \label{topic100}
 \centering
 \begin{tabular}{c}
 \includegraphics[width=\linewidth, clip=true, trim=1.75cm 25cm 6.6cm 1.85cm]{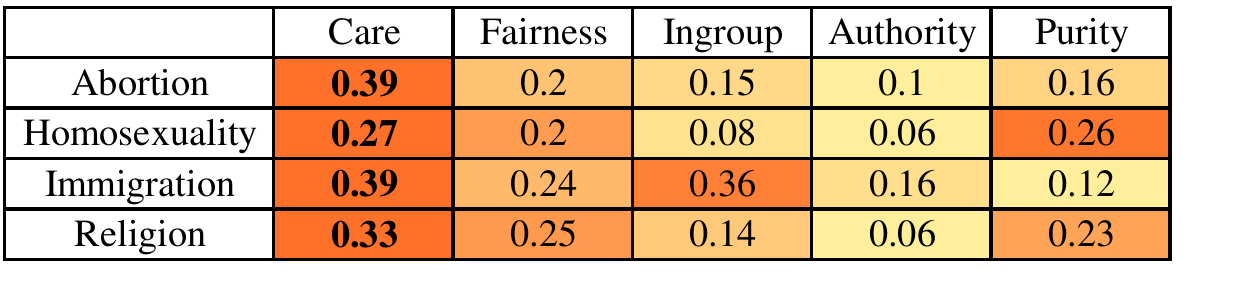} 
 \end{tabular}
 \end{table}
\section{Discussion}
Most social psychology data lie in the bounding box of language, culture, ethnicity, and even environment. 
Social media sites like Facebook and Twitter, however, are useful places for behavioral data mining wherein people without the care of language, ethnicity, and boundaries are free to speak their mind without the need to think of any consequence thereof. 
Therefore, social big data is suitable for observing and measuring human behavior in a natural setting. 
This complements data from social psychology experiments. 

We have analyzed spontaneous moral conversations on Twitter where people wrote about myriad topics such as abortion, homosexuality, immigration, religion, and immorality in general. 
By quantifying moral loadings from tweets, we observed that while the five moral foundations are mutually related, Care is the most dominant foundation and Purity is the most distinctive foundation---at least in online conversations about immorality. 
These findings need to be further tested with a larger size of corpora from various social media sites. 
With the MF dictionary, a simple word counting approach is often used to measure moral charge from texts, but the LSA-based approach used here has some advantages. 
It allows us to capture the meanings of moral words as vectors, and thus there are a wide variety of methods in vector semantics. 
For example, topic modeling is a promising application of NLP to gain new insights into moral psychology. 
Here, we fixed the size of a word-context matrix according to the purpose, but the estimation of the appropriate size is another important issue that future studies need to address.
We are aware that translated versions of the MF dictionary are necessary for cross-cultural comparisons of moral conversations. 
This is also an important future direction.
Although several forthcoming issues remain, the current study demonstrates a new possibility of NLP and social big data for moral psychology by quantifying moral diversity in everyday conversations.

\section*{Acknowledgment}
This work was supported by JSPS KAKENHI grant no.15H03446. We thank M. Karasawa for discussions.

\bibliographystyle{abbrv}
\bibliography{references}
\end{document}